\documentclass{edm_article}
\usepackage{float}
\begin{document}

\title{RAMO: Retrieval-Augmented Generation for Enhancing MOOCs Recommendations\titlenote{This paper underwent a rigorous review process and was officially accepted on May 31, 2024, for presentation at the Educational Data Mining 2024 Workshop: Leveraging Large Language Models for Next Generation Educational Technologies.}}

% Submissions for EDM are double-blind: please do not include any author names or affiliations in the submission. 
% Anonymous authors:
% \numberofauthors{1}
% \author{
% Anonymous\\
%        \affaddr{Anonymous Institution}\\
%        \email{anonymous@anonymous.edu}
% }
%An example of how to include
% multiple authors is below for after the paper has been accepted.


\numberofauthors{2} %  in this sample file, there are a *total*
% % of EIGHT authors. SIX appear on the 'first-page' (for formatting
% % reasons) and the remaining two appear in the \additionalauthors section.
% %
\author{
% % You can go ahead and credit any number of authors here,
% % e.g. one 'row of three' or two rows (consisting of one row of three
% % and a second row of one, two or three).
% %
% % The command \alignauthor (no curly braces needed) should
% % precede each author name, affiliation/snail-mail address and
% % e-mail address. Additionally, tag each line of
% % affiliation/address with \affaddr, and tag the
% % e-mail address with \email.
% %
% % 1st. author
\alignauthor
Jiarui Rao\\
       \affaddr{Carnegie Mellon University}\\
       \affaddr{5000 Forbes Ave }\\
       \affaddr{Pittsburgh, PA 15213}\\
       \email{jiaruira@andrew.cmu.edu}
% 2nd. author
\alignauthor
Jionghao Lin\\
      \affaddr{Carnegie Mellon University}\\
       \affaddr{5000 Forbes Ave }\\
       \affaddr{Pittsburgh, PA 15213}\\
       \email{jionghal@andrew.cmu.edu}
% % 3rd. author
% \alignauthor Lars Th{\o}rv{\"a}ld\titlenote{This author is the
% one who did all the really hard work.}\\
%        \affaddr{The Th{\o}rv{\"a}ld Group}\\
%        \affaddr{1 Th{\o}rv{\"a}ld Circle}\\
%        \affaddr{Hekla, Iceland}\\
%        \email{larst@affiliation.org}
% \and  % use '\and' if you need 'another row' of author names
% % 4th. author
% \alignauthor Lawrence P. Leipuner\\
%        \affaddr{Brookhaven Laboratories}\\
%        \affaddr{Brookhaven National Lab}\\
%        \affaddr{P.O. Box 5000}\\
%        \email{lleipuner@researchlabs.org}
% % 5th. author
% \alignauthor Sean Fogarty\\
%        \affaddr{NASA Ames Research Center}\\
%        \affaddr{Moffett Field}\\
%        \affaddr{California 94035}\\
%        \email{fogartys@amesres.org}
% % 6th. author
% \alignauthor Charles Palmer\\
%        \affaddr{Palmer Research Laboratories}\\
%        \affaddr{8600 Datapoint Drive}\\
%        \affaddr{San Antonio, Texas 78229}\\
%        \email{cpalmer@prl.com}
}

\maketitle

\begin{abstract}

Massive Open Online Courses (MOOCs) have significantly enhanced educational accessibility by offering a wide variety of courses and breaking down traditional barriers related to geography, finance, and time. However, students often face difficulties navigating the vast selection of courses, especially when exploring new fields of study. Driven by this challenge, researchers have been exploring course recommender systems to offer tailored guidance that aligns with individual learning preferences and career aspirations. These systems face particular challenges in effectively addressing the ``cold start'' problem for new users. Recent advancements in recommender systems suggest integrating large language models (LLMs) into the recommendation process to enhance personalized recommendations and address the ``cold start'' problem. Motivated by these advancements, our study introduces RAMO (Retrieval-Augmented Generation for MOOCs), a system specifically designed to overcome the ``cold start'' challenges of traditional course recommender systems. The RAMO system leverages the capabilities of LLMs, along with Retrieval-Augmented Generation (RAG)-facilitated contextual understanding, to provide course recommendations through a conversational interface, aiming to enhance the e-learning experience.

\end{abstract}

 % Replace with your own 3-5 keywords

\section{Introduction}

Massive Open Online Courses (MOOCs) gently facilitate access to learning for a diverse global audience \cite{BATURAY2015427}. By providing an extensive range of courses through an easily accessible online platform, MOOCs not only enhance individual learning and development but also enrich the broader educational community \cite{castillo2015moocs}. However, the diverse categories of courses across disciplines can often overwhelm students when they step into new fields of study \cite{knox2014digital}. Selecting the right courses that align with both personal interests and academic requirements is crucial, as improper choices may lead to wasted time, and resources, and a lack of fulfillment in one's educational journey.

To resolve this, researchers have developed course recommender systems using advanced algorithms to offer tailored guidance that aligns with individual learning preferences \cite{romero2010educational}. Many existing implementations of recommendation systems have demonstrated significant benefits, such as enhancing personalized learning experiences and improving student engagement, as highlighted by a recent study \cite{gulzar2018pcrs}. However, these systems also face critical limitations, particularly the ``\textit{cold start}’’ problem, which occurs when trying to make recommendations for new users with limited historical data \cite{jeevamol2021ontology}. Though previous research proposed a more complex framework\textemdash a novel meta-learning heterogeneous information networks approach \cite{lu2020meta}\textemdash to address the ``\textit{cold start}’’ recommendation issue, the approach faces the challenge of high computational complexity, which is not scalable for large-scale MOOCs platforms.

In response to address the limitations of prior work in recommendation systems, where the recommendations lack sufficient personalization and interaction with users, researchers have proposed integrating large language models (LLMs) into course recommendations \cite{kolb2023potentials}. This approach enhances recommendation accuracy and personalization by leveraging user history and conversational prompts. Recent frameworks like GPT4Rec \cite{li2023gpt4rec} and Chat-Rec \cite{gao2023chatrec} demonstrated the potential of LLMs in improving course alignment with learners' interests and interaction. However, LLMs can sometimes generate misleading or outdated information. To counteract these shortcomings, one possible solution is the integration of Retrieval-Augmented Generation (RAG) with LLMs \cite{lyu2024llmrec}. 

\begin{figure*}[btp]
\Description{Demo of the RAG facilitated course recommender system we developed }  % Alt text
\centering
\includegraphics[width=0.8\textwidth]{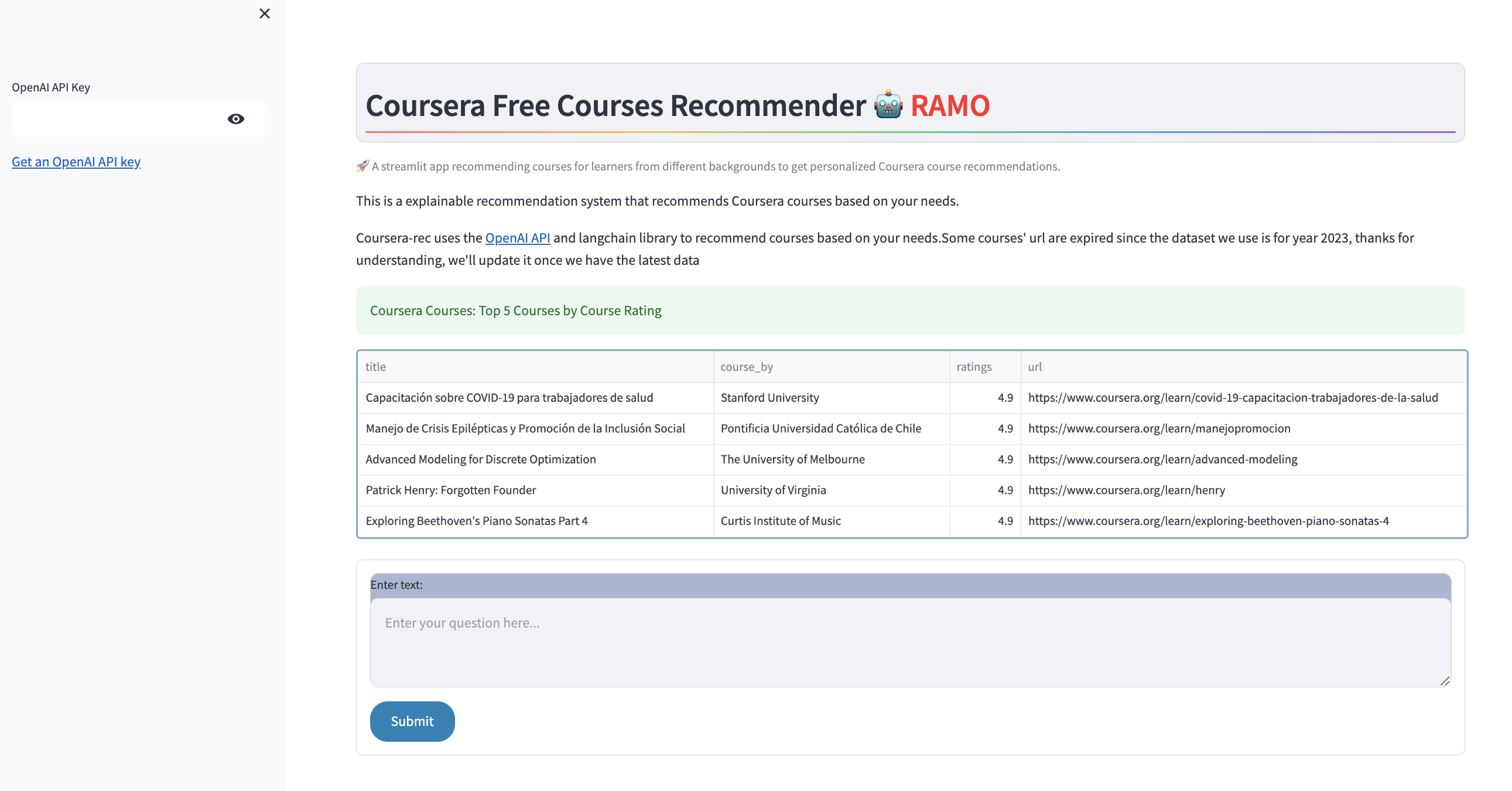} % Adjusts the image to the width of the text
\caption{Interface of the Retrieval-Augmented Generation for MOOCs (RAMO) system}
\label{fig:demo of RAMO}
\end{figure*}

RAG \cite{gao2023retrieval} is a process that optimizes the output of LLMs by extending their robust capabilities to cater specifically to distinct domains or an organization’s internal knowledge base, eliminating the need for retraining the model \cite{feng2024retrieval}. The use of RAG in recommendation systems enhances the adaptability of LLMs, ensuring that recommendations remain current and contextually relevant \cite{lyu2024llmrec}. This advancement paves the way for more precise and targeted course recommendations that adapt to changes in educational content and learner preferences. Despite these improvements, there is a noticeable gap in research specifically focused on using LLMs in course recommender systems, particularly in addressing the ``\textit{cold start}’’ problem where the system lacks a user's profile. Thus, our study aims to investigate the potential of LLMs, particularly those enhanced by RAG, in providing course recommendations tailored to individual user needs. We introduce a course recommender system, \textbf{RAMO} (\textbf{R}etrieval-\textbf{A}ugmented \textbf{G}eneration for \textbf{MO}OCs), which employs a RAG-based LLM model (refer to Figure \ref{fig:demo of RAMO}). RAMO leverages RAG's advantage to improve the quality of course recommendations, addressing and mitigating common issues associated with LLMs especially in ``\textit{cold start}’’ problem.

\section{Related Works}
\subsection{Course Recommender Systems}

Course recommender systems are essential in educational technology, helping students choose courses that align with their interests and academic goals. Many prior studies have employed collaborative filtering methods to build course recommender systems \cite{basilico2004unifying, schafer2007collaborative, koren2021advances}. For instance, Schafer et al. \cite{schafer2007collaborative} proposed a recommender system that suggested courses based on the preferences of similar users. A more recent example by Koren et al. \cite{koren2021advances} developed advanced collaborative filtering techniques to enhance course recommendation accuracy. However, a significant issue arises when recommending courses for new users, as there is no historical data available for these individuals—this is known as the ``\textit{cold start}’’ problem  \cite{wu2024could}. To address this challenge, a recent study by Wu et al. \cite{wu2024could} leveraged large language models (LLMs), which utilize extensive pre-trained knowledge from web datasets, demonstrating potential in overcoming the cold start problem. Despite the advancements in LLMs, their integration into course recommendation systems remains largely unexplored, presenting an opportunity for future research to innovate and improve student course selection processes.

\subsection{Large Language Models in Education}

 Large language models (LLMs) like ChatGPT, trained on extensive datasets, have the ability to generate human-like text and respond to questions with exceptional precision \cite{Hasan2023, ZDNet2024}. Many studies have highlighted the potential of LLMs in educational applications, leveraging their capabilities to enhance various aspects of teaching and learning. For example, Kabir and Lin \cite{kabir2023llm} developed an adaptive practicing system utilizing ChatGPT to generate personalized questions and feedback, demonstrating LLMs' potential in facilitating tailored educational interactions. Researchers investigated multiple GPT models on their ability to generate tailored learning materials and provide instant feedback on student errors, enhancing personalized learning experiences \cite{wang2024large}. Huber et al. \cite{huber2024leveraging} 
 demonstrated the use of LLMs in creating interactive, conversational systems that assist both students and teachers by providing adaptive learning support and resources. Moreover, LLMs are also used in generating automatic feedback for students \cite{dai2023can, daiassessing}, handling sparse learner performance data \cite{zhang20243dg} from intelligent tutoring systems, predicting learning performance \cite{zhang2024predicting}, and supporting tutor training session \cite{lin2024can}.

 \subsection{Retrieval-Augmented Generation in Education}Retrieval-augmented generationn (RAG) has emerged as a significant technique to enhance the effectiveness of educational tools powered by LLMs. For example, a study \cite{levonian2023retrieval} integrated textbook content into LLM prompts via RAG improved the quality of responses in interactive question-answering (QA) scenarios for middle-school math students, and demonstrated that students generally prefer responses generated by RAGs. 
RAG has also been employed in programming education to generate improved feedback for student's completion of coding tasks \cite{jacobs2024leveraging}, by incorporating transcriptions of lecture recordings and using timestamps as meta-information, RAG reduces hallucinations and ensures the use of accurate technical terms.  Moreover, RAG has been utilized to assess novice math tutors' use of social-emotional learning strategies \cite{lin2024improving}, they proved that RAG-enhanced prompts demonstrated more accurate and cost-effective performance compared to other prompting strategies by providing relevant external content. This application highlights the potential of RAG in developing personalized tutor training programs and enhancing the overall effectiveness of tutored learning.

While traditional course recommender systems have laid the groundwork for personalized education, the integration of LLMs and techniques such as RAG offers unprecedented opportunities for enhancing educational experiences. These advanced methods address limitations of earlier approaches and pave the way for more sophisticated and effective educational tools, inspiring us to utilize RAG in developing our course recommender system.

\section{Method}
\subsection{Dataset}
In this study, we utilized the ``Coursera Courses Dataset 2021''\footnote{\url{https://www.kaggle.com/datasets/khusheekapoor/coursera-courses-dataset-2021}} from Kaggle. The dataset, scraped from Coursera's publicly available information in September 2021, contains a variety of courses that feature comprehensive details such as skill requirements, difficulty levels, and direct course links. It provides a robust knowledge base for our RAMO system, enabling it to suggest courses tailored to students' specific skills and educational needs. This dataset effectively supports our objective to enhance accessibility and personalized learning through course recommendations. We first cleaned the dataset to remove meaningless symbols and duplicate rows, and it has 3,342 non-duplicate courses in total after data-cleaning, with 6 columns:

\begin{itemize}
    \item \textbf{Course Name}: The title of the course.
    \item \textbf{University}: The institution offering the course.
    \item \textbf{Difficulty Level}: The level of complexity of the course content.
    \item \textbf{Rating}: The average rating given by learners.
    \item \textbf{URL}: The web address where the course can be accessed.
    \item \textbf{Description}: A brief overview of what the course covers.
    \item \textbf{Skills}: The specific abilities or knowledge areas that the course aims to develop.
\end{itemize}

\subsection{Recommendation System Design}

\subsubsection{Prompt Design}
The ``cold start'' problem, where systems lack user historical data, is a significant challenge in recommendation systems. Both traditional course recommender algorithums like content-based and collaborative-filtering algorithms and LLM-based system recommendation systems struggle with this issue. However, our RAG-based solution addresses this by using a `prompt template' in the back-end. This template guides RAMO to generate relevant responses even when no user-specific data is available, as detailed in Table \ref{tab:interaction_prompts}. The RAMO system can provide meaningful recommendations from the outset, unlike non-RAG-based recommender systems, which lack a retrieval process and prompt-based customization. The prompt to our retriever (i.e., to retrieve the relevant docs from the databases) is called the `prompt template', which is shown in Table \ref{tab:interaction_prompts}. The prompt to our generator is composed with three parts: 1) \texttt{User Question}, 2) \texttt{Prompt Template}, and 3) \texttt{Search Results} (the context of the retrieved relevant documents). We also added the uplifting adverb `\textit{fantastic}' to the prompt template, to elevate it with Emotional Intelligence since ChatGPT is designed to recognize patterns in language, including those associated with emotions \cite{vinay2024emotional}.

\begin{table}[h!]
\centering
\caption{Overview of interaction prompt structure}
\label{tab:interaction_prompts}

\begin{tabular}{|p{8cm}|} % Adjust the width as needed to fit your page layout
\hline
\textbf{Prompt Template} \\ 
\textit{You are a fantastic Coursera course recommender. Use the following pieces of context to answer the question and recommend relevant courses to the user.
If the user doesn't specify their requirements, you can just recommend some courses that are most popular in the system based on their ratings and difficulty levels. You only need to provide the course title to the user.
Also, please pay attention to how many courses the user wants you to recommend.
If you don't know the answer, just say ``I don't know''.} \\ \hline
\textbf{Context} \\ 
Retrieved course data \\ \hline
\textbf{User Question} \\ 
User's specific question to the generator \\ \hline
\end{tabular}
\end{table}

\subsubsection{Integration of RAG approach}
As shown in Table \ref{tab:cost of models} below, we employed several LLMs to build our course recommender system. We provide a list of the LLM models we used, along with details on their associated costs and token limits.  The token limit refers to the maximum number of tokens (a token represents about 3/4 of a word or four characters, according to Open AI \cite{netdocuments}) that the model can process in a single input. While some models, like Llama 2 and Llama 3, are free to use on small-scale dataset, due to their open-source nature, others may incur costs based on usage or subscription plans \cite{Meta2024}.

\begin{table}[h!]
\centering
\renewcommand{\arraystretch}{1.2}  % Provide more space between table rows, if you prefer
\caption{Cost and token limit of models we used}
\label{tab:cost of models}
\begin{tabular}{ccl} \hline
LLM Model& Output Cost&Token Limit\\ \hline
\texttt{GPT-3.5 Turbo} & $0.50$ per 1M tokens & 4,096 tokens\\
\texttt{GPT-4} & $30.00$ per 1M tokens & 8,192 tokens \\ 
\texttt{Llama-2} & Free & 4,096 tokens \\ 
\texttt{Llama-3} & Free & 8,000 tokens \\ 

\hline\end{tabular}
\end{table}

\begin{figure*}[h!]
\Description{Workflow of the RAG facilitated course recommendation system we proposed}
\centering
\includegraphics[width=2.00\columnwidth]{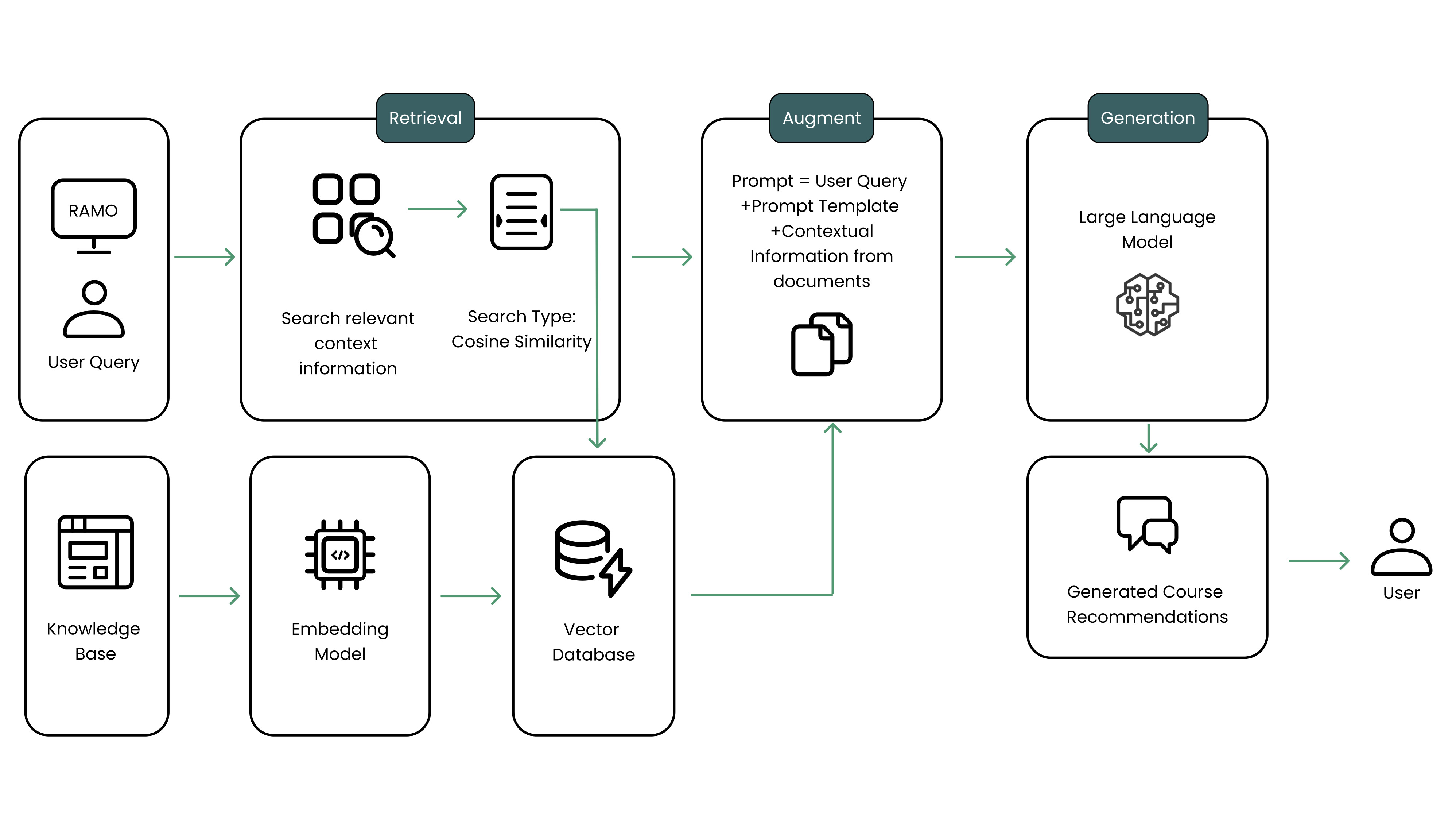}
\caption{Workflow for the RAMO System}
\label{fig:diagram of RAMO}
\end{figure*}

We then leveraged the RAG approach to enhance the system's understanding of the user context. As shown in Figure \ref{fig: diagram of RAMO}, RAG consists of two primary components: the \textit{retriever} and the \textit{generator}. The retriever aims to enhance the prompt templates, which `augment' the retrieval process, tailoring it to specific user queries. The knowledge base used for the retrieval process can contain any format of course data (e.g., csv, pdf, and json), providing a flexible and rich source of information for generating responses and we used the largest MOOC platform\textemdash coursera's course dataset in csv format as the knowledge base. The dataset was transformed into text embeddings and stored in the vector database. These embeddings were then used to find high-quality, relevant information, which was incorporated into the prompt for the generator. Here we use OpenAI embedding model (\texttt{text-embedding-ada-002} \cite{OpenAI2024}) to tokenize the course data and store the embeddings in vector store, considering its advantage over BERT (Bidirectional Encoder Representations from Transformers) \cite{devlin2019bert}, while OpenAI embeddings \cite{OpenAI2024} offer better generalization and contextual understanding \cite{Moradiya2023}, making them more suitable for diverse educational content. The generator is powered by LLMs, which generate the textual contents based on the engineered prompts. To facilitate user's interaction with the system, we make the recommendation process to be completed via conversational manner. 

The interface of our recommender system is shown in Figure \ref{fig:demo of RAMO}, where we listed 5 default courses based on their ratings in the dataset on the web page to make it more user-friendly.  As for the implementation of the system, we use \texttt{GPT-3.5 Turbo}, selected for its robust integration with the \texttt{LangChain} \cite{topsakal2023creating} framework\textemdash a platform designed to streamline the implementation of language models in application-specific contexts. This setup allows the system to dynamically retrieve relevant documents and generate responses tailored to user inputs, as illustrated in the workflow in Figure \ref{fig:diagram of RAMO}.

\subsubsection{Comparative Analysis}
To evaluate the performance of our system, we conducted a series of tests by providing different prompts representing various user needs to RAMO. This allowed us to explore its ability to deliver course recommendations based on the outputs generated in response to varied user prompts.

\textbf{LLM vs. Non-LLM.} We explored both the relevance of the recommended courses to the user's interests and responding time (the time it takes to generate a response) of the LLM-based recommender system compared to non-LLM course recommender systems (e.g., course recommender system using collaborative filtering and content-based approaches), focusing particularly on their ability to address the ``\textit{cold start}'' problem. This problem occurs when the user lacks specific requirements on what skills they want to learn, and the system lacks data on the new user.

\textbf{LLM vs. LLM with RAG.} We further examined the performance of a standard LLM recommender system (without RAG and without using a dataset as a knowledge base) versus an RAG-enhanced LLM recommender system by testing different prompt templates for the retriever and various user queries for the generator to ascertain improvements in system performance and recommendation personalization.

To explore the performance of our course recommender system, we focused on comparing the relevance of the recommended courses to different prompts by varying prompt templates and user-specific requirements.

\section{Results}
\subsection{LLM vs. Non-LLM}

\begin{figure*}[h!]
\Description{LLM output Vs Rag-Based Output for New User who want to learn something on coursera}
\centering
\includegraphics[width=2.00\columnwidth]{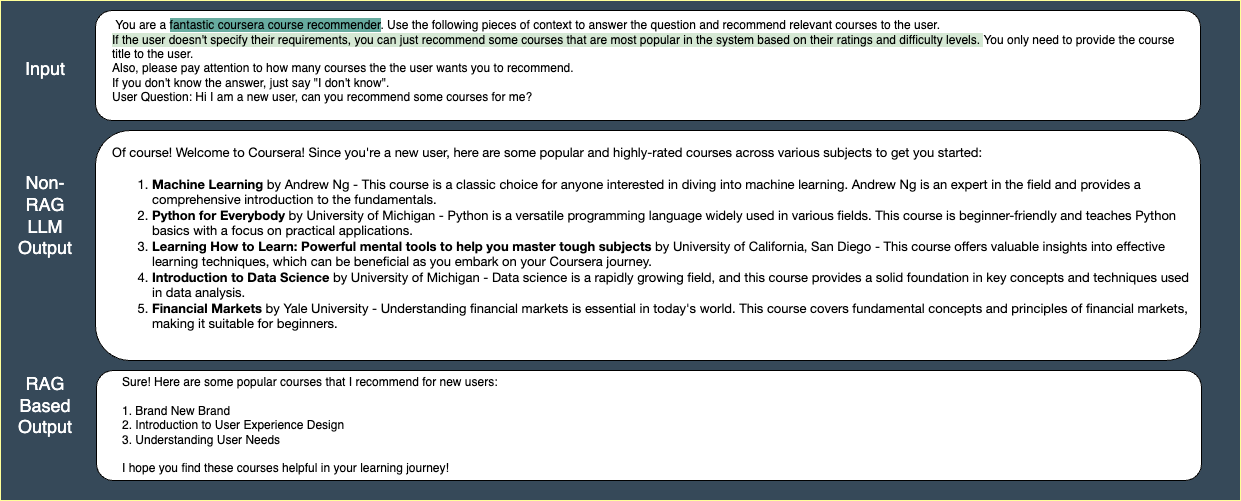}
\caption{Sample output for a cold-start question on LLM vs RAG-LLM system}
\label{fig:cold start for LLM vs RAG}
\end{figure*}

We compared RAMO with a traditional course recommendation system built by the content-based and collaborative filtering using the same dataset\footnote{\url{https://www.kaggle.com/code/sagarbapodara/coursera-course-recommendation-system-webapp}}. During this comparison, we focused on the ``\textit{cold start}'' problem. The ``\textit{cold start}'' problem is especially pertinent in the context of an e-learning platform for tutor training, such as tutor training platform \cite{lin2023personalized}. When new tutors join the platform, they are encouraged to complete various training courses to enhance their tutoring skills. Given the wide range of courses available, new tutors may feel overwhelmed when deciding where to begin their learning journey. In such scenarios, they may ask general questions such as, ``\textit{What can I learn today since I am a new tutor onboarding to this platform?}'' They do not have prior course completions or specific learning preferences logged in the system, making it challenging for the recommendation system to personalize suggestions based on historical data. When prompted with ``\textit{I am a new user}'', the traditional recommender system failed to generate a recommendation because its algorithm relies on the cosine similarity of the descriptive texts of the user's desired learning topic and the database items, and there are no courses with similar title or description as the phrase `\textit{new user}'. In contrast, both our standard LLM and the RAG-enhanced LLM system can provide relevant course suggestions for the new user, with the LLM offering more detailed descriptions based on its internal knowledge base and RAG offering more customized outputs based on its external knowledge base and the prompt template we designed. The comparative results for both the standard and RAG-based recommender systems are displayed in Figure \ref{fig:cold start for LLM vs RAG}.

Regarding system performance, the traditional system typically took about 0.02 seconds longer than RAMO to generate responses according to the same user interest\textemdash a certain topic the user wants to learn, and this delay increased with the complexity of the user's input regarding relevant skills.

\subsection{LLM vs. LLM with RAG}

\begin{figure}[h!]
\Description{User Entry and the Output part of the system we developed}
\centering
\includegraphics[width=1\columnwidth]{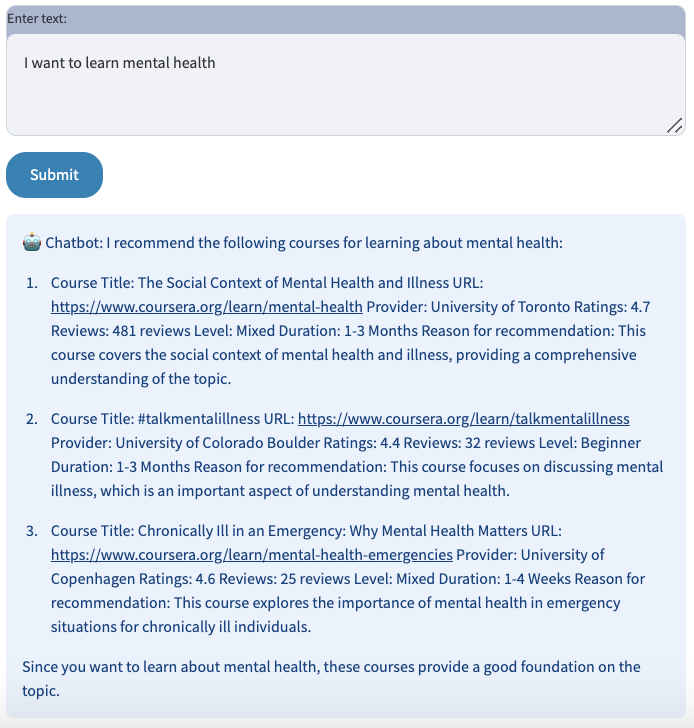}
\caption{Output for a specific user question}
\label{fig: output of a question}
\end{figure}

To explore how well our LLMs can provide personalized course recommendations, we used prompts that specified a particular skill to be learned. The non-RAG LLM (based on GPT-3.5) delivered detailed suggestions for relevant courses available on Coursera, utilizing its internal database of courses. In contrast, the recommendations from the RAG-enhanced LLM varied according to the specific prompt template used by the retriever. This adaptability allows developers to tailor the quantity and detail of the courses recommended, showcasing the flexibility of the RAG approach. The user interface and the outcomes for a query focused on learning a specific skill are illustrated in Figure \ref{fig: output of a question}.

\begin{figure}[h!]
\Description{Different User Questions and Responses of the Generator }  % Alt text
\centering
\includegraphics[width=0.98\columnwidth]{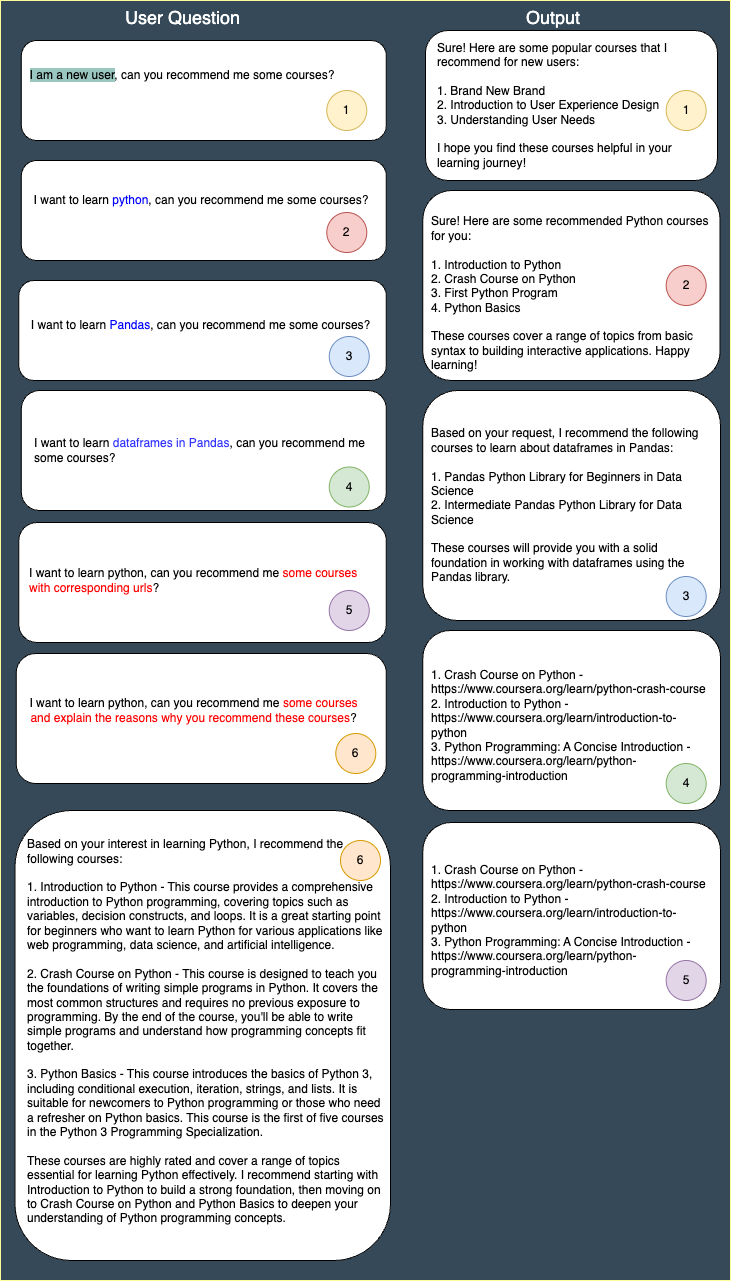} % Adjusts the image to the width of the text
\caption{User questions and related outputs}
\label{fig: different user questions}
\end{figure}

We modified the retrieval prompts and generation queries to test the adaptability of our recommendation system. First, we conducted tests on various user queries using the same prompt template to compare the variations in output. The first module in Figure \ref{fig: different user questions} illustrates the system's response to a ``\textit{cold start}'' problem, while modules 2 through 6 demonstrate how the output varies based on user questions about the number of courses recommended and the level of detail provided, such as reasons for recommendations, URLs, and other specifics. For example, when user asks question like ``\textit{I want to learn python, can you recommend me some courses?}'', RAMO can give the output to the user: ``\textit{Sure! Here are some recommended Python courses for you:
1. Introduction to Python
2. Crash Course on Python
3. First Python Program
4. Python Basics
These courses cover a range of topics from basic syntax to building interactive applications. Happy learning!}'' When the user changes their mind and decides to learn about another topic, RAMO can give relevant recommendations. The outputs consistently matched the user requirements in relevance, successfully retrieving the pertinent courses from the Coursera dataset, more examples could be found at Figure \ref{fig: different user questions}.

\begin{figure}[h!]
\Description{Different Prompt Templates and Responses of the Generator }  % Alt text
\centering
\includegraphics[width=0.98\columnwidth]{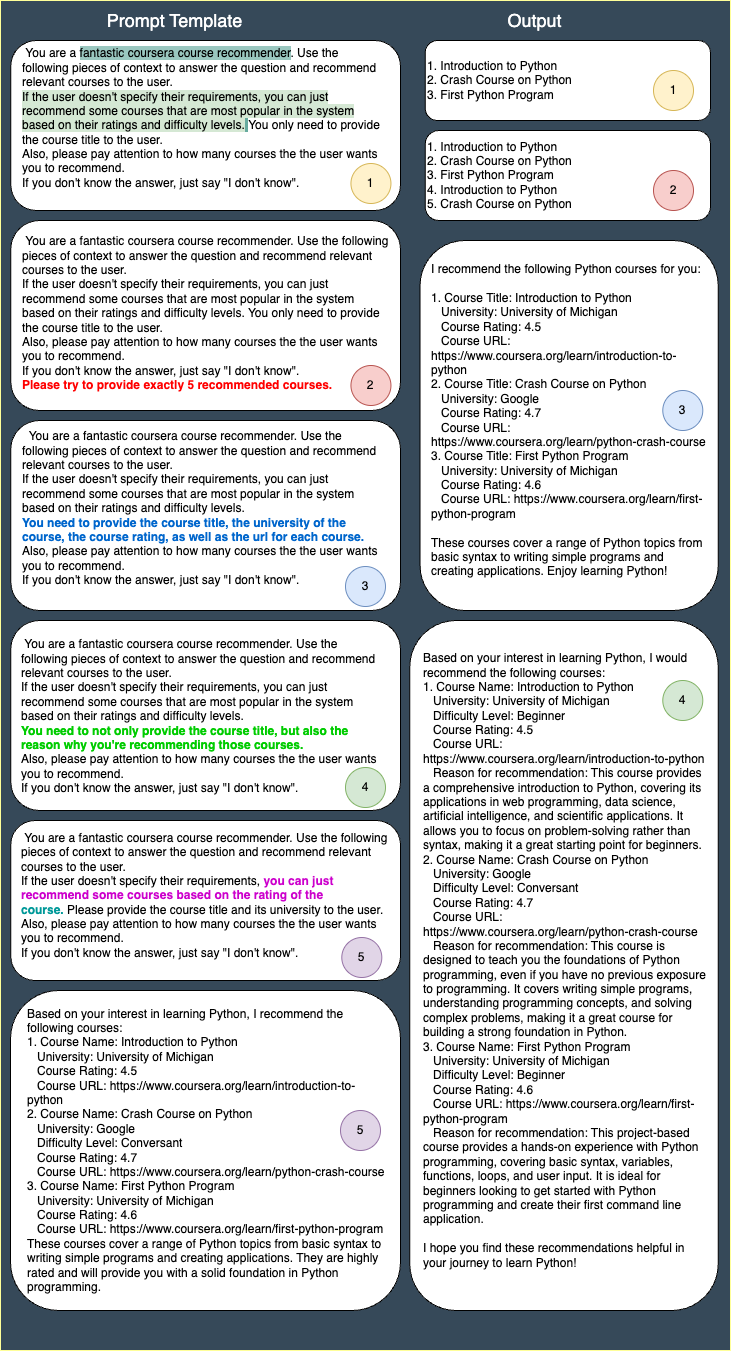} % Adjusts the image to the width of the text
\caption{Prompt templates and related outputs}
\label{fig: different prompt templates}
\end{figure}

We also utilized different retrieval prompt templates to explore how the output varies based on different prompts.  Specifically, we used the same user question ``\textit{I want to learn python}'', and altered the prompt templates to specify the number of recommended courses and the level of detail provided in the output, ranging from mere course titles to comprehensive descriptions that include titles, URLs, and rationales for each recommendation. The variations in the prompt templates and their corresponding outputs are illustrated in Figure \ref{fig: different prompt templates}.  Here, red lines highlight changes in the number of courses recommended, blue lines detail the content of the courses—such as the inclusion of reasons for recommendations or just the course titles, ratings, and URLs\textemdash while green highlights how we addressed the ``\textit{cold-start}'' problem, resulting in recommendations of the three most popular (based on course ratings) and easiest courses (based on its difficulty level), as depicted in the output module labeled 1 in Figure \ref{fig: different prompt templates}. The generated response in response to varied prompts underscores the system's robustness; for instance, when the template specifies ``\textit{recommend three courses at a time}'', the output consistently includes exactly three courses. Similarly, if the prompt contains `course URLs and titles', the system reliably appends this information to each recommended course, ensuring that the output meticulously adheres to the specified criteria.

\section{Conclusions}
In this study, we have demonstrated the application of LLMs as course recommender systems, particularly within MOOCs. Our findings confirm the potential of LLMs to deliver personalized course recommendations based on user's different requirements. We initially compared four LLMs, including GPT-3.5 Turbo and GPT-4. Ultimately, we selected GPT-3.5 as the back-end model for the RAMO system due to its comparable performance to GPT-4 at a lower cost. Although the Llama models are free to access, we found that the GPT models were significantly faster. Specifically, GPT-3.5 had an approximate response time of 3 seconds, whereas Llama 2 and Llama 3 took approximately 5 minutes and 8 minutes, respectively. Furthermore, the integration of RAG has enhanced the quality of recommendation outputs, as evidenced by the generated responses based on various user prompts, which are highly related to user's needs and all came from the knowledge base. Additionally, our system supports conversational interaction with users, which could be seamlessly integrated into numerous online educational platforms. Our use of open-source LLMs (e.g. Llama 2 and Llama 3 \cite{Meta2024}) has also been validated, proving to be a cost-effective approach for broader deployment.

\textbf{Limitations}\newline
As this study is ongoing, we have not yet conducted comprehensive evaluations of our recommender systems, including human evaluations or user studies. This is primarily due to the nascent stage of our research. Moreover, while many research projects on recommendation systems employ benchmarks to evaluate system adaptability, our study currently lacks such benchmarks because we do not possess a test dataset. The Coursera dataset we utilized includes only course data, lacking user profiles which are essential for evaluating the effectiveness of recommender systems across different time periods. If we had access to user data, including users' past course learning histories and their preferences, we could integrate this information with the course data to enhance our retrieval process. This integration would allow us to personalize recommendations more effectively, tailoring course suggestions to individual learning patterns and preferences. Incorporating detailed user data would enable RAMO to provide more accurate and relevant recommendations, improving user satisfaction and engagement. It would also allow for longitudinal studies to track how users' interactions with the system evolve over time and how well the recommendations align with their long-term learning goals.\newline
\newline
\textbf{Future Work}\newline
We plan to undertake several further steps to advance our research. \textit{Firstly}, we aim to conduct thorough evaluations and tests to validate the efficacy and reliability of our recommender systems. This will involve integrating user studies and utilizing real user data once our systems are deployed on our e-learning platform. Such measures will enable us to robustly measure performance and refine our approach. \textit{Secondly}, we will focus on enhancing system performance, considering scalability and the potential to expand our technology to encompass a broader range of educational tools and platforms. These efforts will ensure that our recommender systems not only meet current educational needs but also adapt to future demands and technological advancements. \textit{Thirdly}, we could deploy RAMO on our own e-learning platform, and then have the opportunity to gather comprehensive user data and utilize our own course dataset rather than Coursera's. This deployment would allow us to conduct extensive testing and validation, further proving the eligibility and effectiveness of the LLM for recommending courses. With access to real-time user data, we could continuously refine our algorithms, making the system more adaptive and responsive to users' evolving needs.

% \textbf{Experiment Design}\newline 
To evaluate the effectiveness of our LLM-based course recommendation system, we plan to conduct a comprehensive experiment that includes quantitative metrics, user studies, and personalization improvements. Our experiment aims to assess both the relevancy of the recommendations and the satisfaction of the users with the recommended courses.

% \textbf{Quantitative Metrics}\newline 
We will utilize several quantitative metrics to evaluate the performance of the recommendation system. Key metrics include post-test performance, measured by the improvement in students' scores from pre-test to post-test after tutoring sessions, and course completion rate, which compares the rate of course completion between students who follow the system's recommendations and those who do not. Additionally, engagement rate will be tracked by monitoring whether students continue engaging with the lesson without dropping out midway. User satisfaction will also be assessed through feedback collected after each lesson via a thumbs-up or thumbs-down system and detailed surveys. To gather qualitative insights into the system's effectiveness and user experience, we will conduct user studies. These will involve satisfaction surveys completed by students following each lesson to gauge their satisfaction with the course content and the relevance of the recommendations, as well as focus group discussions to explore students' experiences in more depth and gather suggestions for improvement.

%\end{document}  % This is where a 'short' article might terminate

%ACKNOWLEDGMENTS are optional
\section{Acknowledgments}
We extend our sincere gratitude to Chenfei Lou, a current software engineer at X (former twitter), for his invaluable guidance in developing our demo. We also thank Sandy Zhao, a current master’s student in the CMU METALs program, for her excellent assistance in generating the wonderful diagram. Additionally, we appreciate Yuting Wang, an undergraduate student at CMU, for her help in refining the design in this paper.

\bibliographystyle{abbrv}
\bibliography{EDM_Article_Submission}  % sigproc.bib is the name of the Bibliography in this case
% \balancecolumns
\appendix
%Appendix A
\section{DATASET WE USE}
The 2021 coursera dataset we use is available at \url{https://www.k
aggle.com/datasets/khusheekapoor/coursera-courses-dataset-2021}.
\section{Link of the RAG system code}
The code could be run at:\url{https://colab.research.google.com/drive/1wLwM5QphDoIctW9_EZt26D6RIpSmfaCD?usp=sharing}, including the data preprocessing and the RAG process.
\section{Link of the Chatbot DEMO}
The demo of the chatbot could be accessed at:\url{https://huggingface.co/
spaces/dinosaur-organization/coursera-recommendation#/}, you need to first obtain your Openai Api and enter it on the left side.
\section{Link of the Medium Blog inspired us of the demo design}
We got inspired by a blog published on medium on the designing of a book recommendation chatbot, the article is available at:\url{https://medium.com/data-and-beyond/data-science-and-machine-learning-books-recommendation-chatbot-83757cbb92f7#/
}

% \section{Interaction Prompt structure}

\balancecolumns
% That's all folks!
\end{document}